\documentclass[twocolumn,showpacs,preprintnumbers]{revtex4}
\usepackage{amsfonts}
\usepackage{amssymb}
\usepackage{amsmath}
\usepackage{graphicx}
\usepackage{epsfig}

\begin{document}

\title{Exact results for the criticality of quench dynamics in quantum Ising
models}
\author{Ying Li}
\author{M.X. Huo}
\author{Z. Song}
\email{songtc@nankai.edu.cn}
\affiliation{Department of Physics, Nankai University, Tianjin 300071, China}

\begin{abstract}
Based on the obtained exact results we systematically study the quench
dynamics of a one-dimensional spin-1/2 transverse field Ising model with
zero- and finite-temperature initial states. We focus on the magnetization
of the system after a sudden change of the external field and a coherent
time-evolution process. With a zero-temperature initial state, the quench
magnetic susceptibility as a function of the initial field strength exhibits
strongly similar scaling behaviors to those of the static magnetic
susceptibility, and the quench magnetic susceptibility as a function of the
final field strength shows a discontinuity at the quantum critical point.
This discontinuity remains robust and always occurs at the quantum critical
point even for the case of finite-temperature initial systems, which
indicates a great advantage of employing quench dynamics to study quantum
phase transitions.
\end{abstract}

\maketitle

\section{Introduction}

The study of quantum phase transitions (QPTs) is a fascinating topic in
condensed matter physics and quantum information science. QPTs distinctively
from temperature-driven critical phenomena occur due to the competition
between different parameters describing the interactions of the system, and
QPTs occur only at zero temperature \cite{SachdevBook}. In the second-order
QPT, the ground state undergoes qualitative changes when an external
parameter passes through the quantum critical points (QCPs). The QCPs are
characterized by the divergence in the correlation length, which leads to
the critical scaling behaviors governed by a class of universal exponents
\cite{ScalingBook}. In principle, experimental observations of QPTs could be
achieved at zero temperature \cite{SachdevBook}. However, in practice, it is
difficult to realize since cooling matter to zero temperature is impossible
in any experiment. Hence, recently, the finite-temperature properties of QPT
systems begin to attract more attention \cite{Thermal}. In their results,
the quantum criticality can persist up to a surprisingly high temperature.
However, the critical behavior at finite temperature is not exact but a
remanent of that at zero temperature. Some divergent physical quantities at
the QCP, e.g., the magnetic susceptibility in the one-dimensional spin-1/2
transverse field Ising model (TFIM), become convergent and shifted away from
the QCP in the parameter space at finite temperature \cite{ScalingBook}.

On the other hand, besides the ground state, the dynamic properties of QPT
systems also cause a lot of interests \cite%
{Zurek,Dziarmaga,Sengupta,Anatoli,Greiner,Sadler,Sachdev,Silva,Zhu,QS}.
Actually, the zero-temperature static properties are naturally and
intimately linked to the dynamic process, in which many excited states are
involved. As we know, the scaling exponents are dependent on the effective
dimensionality, which is the sum of the dimension and the dynamic exponent
\cite{SachdevBook,ScalingBook}. For the adiabatic approach of the QPT system
to the QCP, the variation of parameter needs to slow down to infinitesimal
\cite{Anatoli}. So the ground state at the QCP could not be achieved
adiabatically. Differently, quench dynamics can pass through the QCP without
any restriction. In experiments, the quench dynamics of the Bose-Hubbard
model have been performed with ultracold bosonic atoms in optical lattices
\cite{Greiner}. Quench dynamics and their critical behaviors in the TFIM are
also investigated theoretically \cite{Sachdev,Silva}. However, none of them
discussed the effect of temperature on these critical behaviors. The quench
dynamics relate to all of the eigenstates more than just the ground state,
so the quench dynamic properties at the QCP should not be as sensitive to
temperature as those of the ground state.

With these motivations, in this paper, we study the quench dynamics of the
TFIM with zero-temperature and finite-temperature initial states. Initially,
the system is modulated to be with a transverse field strength $\lambda _{i}$
and temperature $T$. The field strength is suddenly changed to $\lambda _{f}$
and the system begins to evolve coherently. Here, we use the magnetization
per spin in the transverse field direction $\mathfrak{m}(\lambda
_{i},\lambda _{f},T;t)$ to characterize the state at time $t$. After a long
enough time evolution, the off-diagonal contributions to $\mathfrak{m}%
(\lambda _{i},\lambda _{f},T;t)$ are cancelled with each other
\cite{Dephasing}, and $\mathfrak{m}(\lambda _{i},\lambda _{f},T;t)$
achieves a steady value finally. The final asymptotic magnetization
is
\begin{equation}
m_{q}(\lambda _{i},\lambda _{f},T)=\mathfrak{m}(\lambda _{i},\lambda
_{f},T;\infty ),
\end{equation}%
which is called the quench magnetization in this paper. To make the
above formula more easily tractable mathematically, we write it as
\cite{Dephasing}
\begin{equation}
m_{q}(\lambda _{i},\lambda _{f},T)=\lim_{\tau
\rightarrow \infty }\int_{0}^{\tau }\frac{dt}{\tau
}\mathfrak{m}(\lambda _{i},\lambda _{f},T;t).
\end{equation}

We will show that $m_{q}$\ as well as its derivatives, the magnetic
susceptibilities $\chi _{i}(\lambda _{i},\lambda _{f},T)=$ $\partial
m_{q}/\partial \lambda _{i}$ and $\chi _{f}(\lambda _{i},\lambda _{f},T)=$ $%
\partial m_{q}/\partial \lambda _{f}$, exhibit critical behaviors when $%
\lambda _{i}$ or $\lambda _{f}$ passes through the QCP. For a
zero-temperature initial state, the susceptibility $\chi _{i}$ diverges
logarithmically and exhibits scaling behaviors\ when $\lambda _{i}$ is in
the vicinity of the QCP. For an initial state with temperature $T$, the
susceptibility $\chi _{f}$ experiences a jump $\delta \chi _{f}=\tanh
(\Delta _{i}/2T)$ when $\lambda _{f}$ is at the QCP, where $\Delta _{i}$ is
the energy gap between the ground and first excited states of the initial
system. The jump $\delta \chi _{f}$ achieves maximum when $T$ turns to zero.
Since this jump is not sensitive to $T$ and the operations are not
restricted by adiabatic conditions, the quench dynamic process has the
advantage to study critical behaviors of QPT systems without the rigorous
restriction of zero temperature.

\section{Quench magnetization}

The TFIM is a famous model for studying the second-order QPTs. It is exactly
solvable and useful for verifying many new concepts and methods. The
Hamiltonian of the TFIM is
\begin{equation}
H\left( \lambda \right) =-\sum\limits_{j=1}^{N}(\sigma _{j}^{z}\sigma
_{j+1}^{z}+\lambda \sigma _{j}^{x}),
\end{equation}%
where $\sigma ^{\alpha }$ is the Pauli matrix ($\alpha =x,y,z$) and $N$ is
the number of sites. The QCP of this system is at $\lambda =1$. Initially,
the strength of the transverse field is $\lambda _{i}$ and the system is in
a thermal state $\rho (\lambda _{i},T;0)=Z^{-1}\exp [-H(\lambda _{i})/T]$
with temperature $T$, where $Z=\mathrm{Tr}\exp [-H(\lambda _{i})/T]$. Then
the transverse field is suddenly changed to $\lambda _{f}$ and the system
begins to evolve coherently driven by $H(\lambda _{f})$. The magnetization
per spin in the transverse field direction is
\begin{equation}
\mathfrak{m}(\lambda _{i},\lambda _{f},T;t)=\mathrm{Tr}\Sigma ^{x}\rho
(\lambda _{i},T;t),
\end{equation}%
where $\Sigma ^{x}=1/N\sum_{j=1}^{N}\sigma _{j}^{x}$ and $\rho (\lambda
_{i},T;t)=e^{-iH(\lambda _{f})t}\rho (\lambda _{i},T;0)e^{iH(\lambda _{f})t}$%
\ is the state at time $t$. After a long enough time evolution, $\mathfrak{m}
$ approaches the steady quench magnetization
\begin{equation}
m_{q}(\lambda _{i},\lambda _{f},T)=\mathrm{Tr}\widehat{m}_{q}(\lambda
_{f})\rho (\lambda _{i},T;0),  \label{mq}
\end{equation}%
where the quench magnetization operator $\widehat{m}_{q}$ is
\begin{equation}
\widehat{m}_{q}(\lambda _{f})=\sum_{n}\left\vert n(\lambda
_{f})\right\rangle \left\langle n(\lambda _{f})\right\vert \Sigma
^{x}\left\vert n(\lambda _{f})\right\rangle \left\langle n(\lambda
_{f})\right\vert .
\end{equation}%
Here, $\{\left\vert n(\lambda )\right\rangle \}$ is the complete set of
eigenstates of $H(\lambda )$, i.e., $H(\lambda )\left\vert n(\lambda
)\right\rangle =E_{n}(\lambda )\left\vert n(\lambda )\right\rangle $, and $%
n=0$ corresponds to the ground state.

Using the exact solution of TFIM \cite{SachdevBook,Dziarmaga,Pfeuty} and
taking the thermodynamic limit $N\rightarrow \infty $, we get the quench
magnetization as
\begin{eqnarray}
m_{q}(\lambda _{i},\lambda _{f},T) &=&\int_{0}^{2\pi }\frac{dk}{2\pi }\tanh
[\epsilon (\lambda _{i},k)/2T]  \label{In} \\
&&\times \cos \theta (\lambda _{f},k)\cos [\theta (\lambda _{f},k)-\theta
(\lambda _{i},k)],  \notag
\end{eqnarray}%
where the function $\theta (\lambda ,k)$ is defined as $\cos \theta (\lambda
,k)=2(\lambda -\cos k)/\epsilon ,$ $\sin \theta (\lambda ,k)=2\sin
k/\epsilon $ and $\epsilon (\lambda ,k)=$ $2\sqrt{1+\lambda ^{2}-2\lambda
\cos k}$. The quench magnetization is dependent on two systems, the initial
system with the transverse field strength $\lambda _{i}$ and the final
system with $\lambda _{f}$. In the following, we will consider two cases, $%
\lambda _{i}\sim 1$ and $\lambda _{f}\sim 1$, respectively.

%%%%%%%%%%%%%%%%%%%%%%%%%%%%%%%%%%%%%%%%%%%%%%%%%%%%%%%%%%%%%%%%%%%%%%%%%%%%%%%

\begin{figure}[tbp]
\includegraphics[bb=15 310 545 735, width=4 cm]{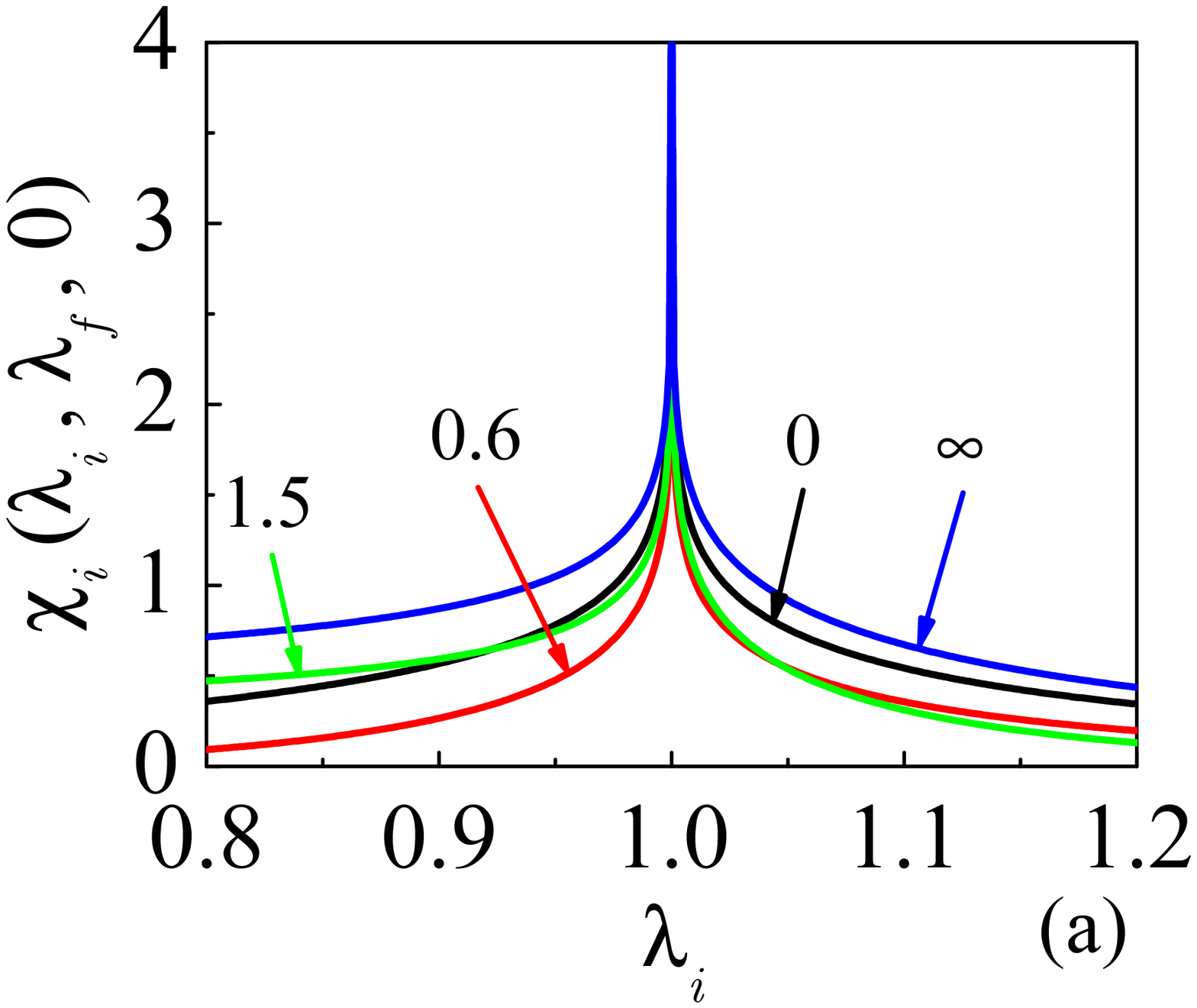} %
\includegraphics[bb=15 310 545 735, width=4 cm]{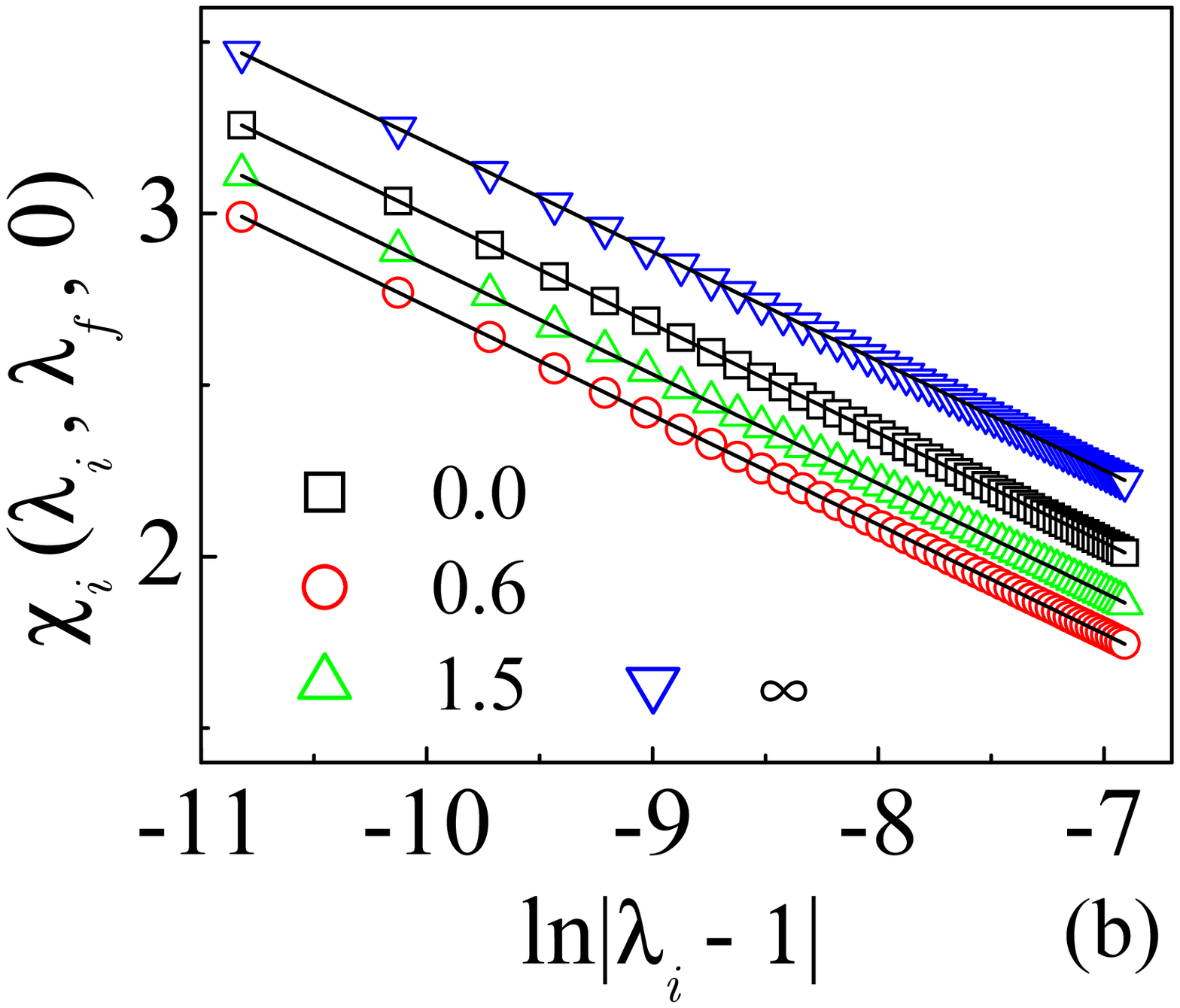} %
\includegraphics[bb=15 310 545 735, width=4 cm]{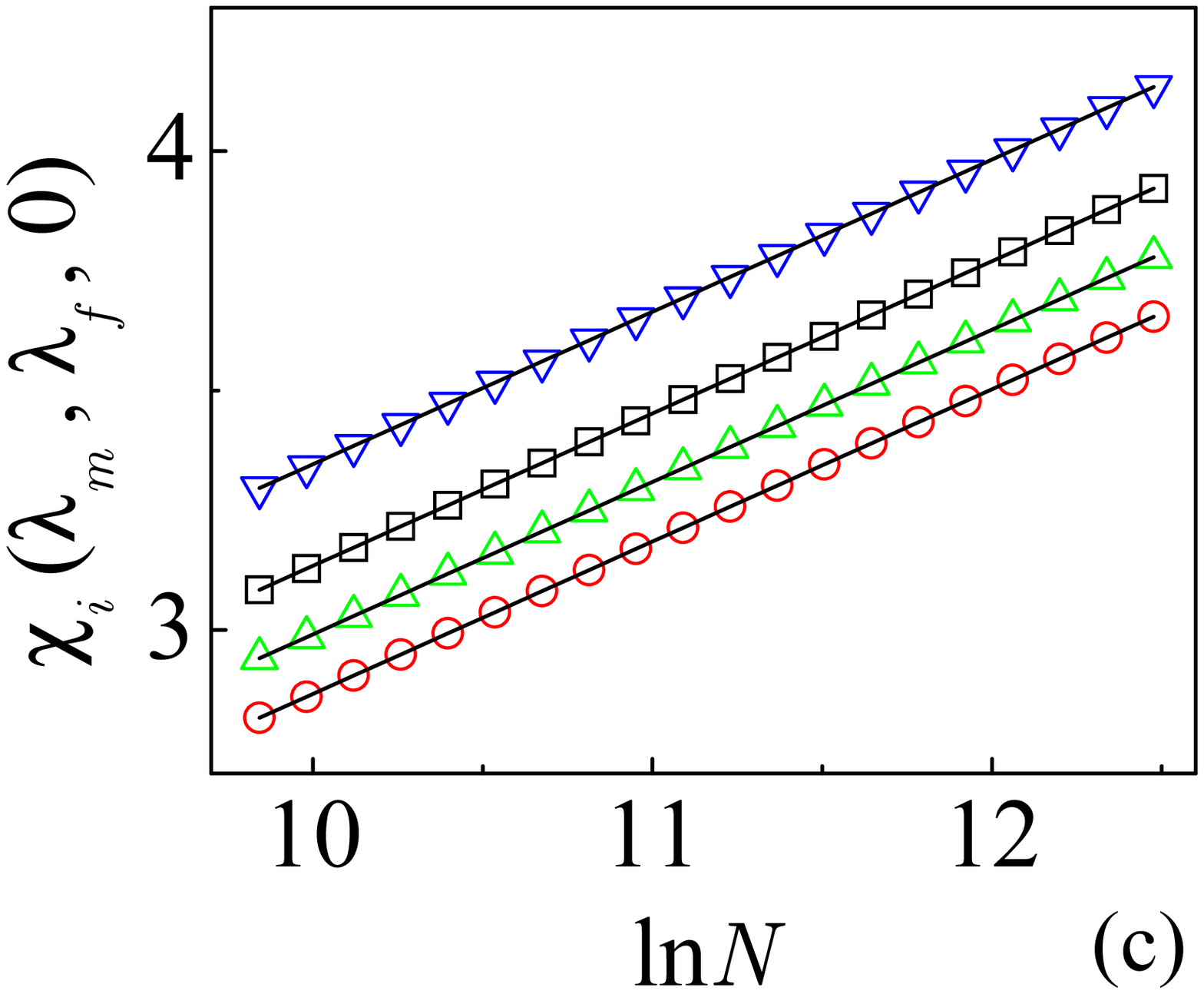} %
\includegraphics[bb=15 310 545 735, width=4 cm]{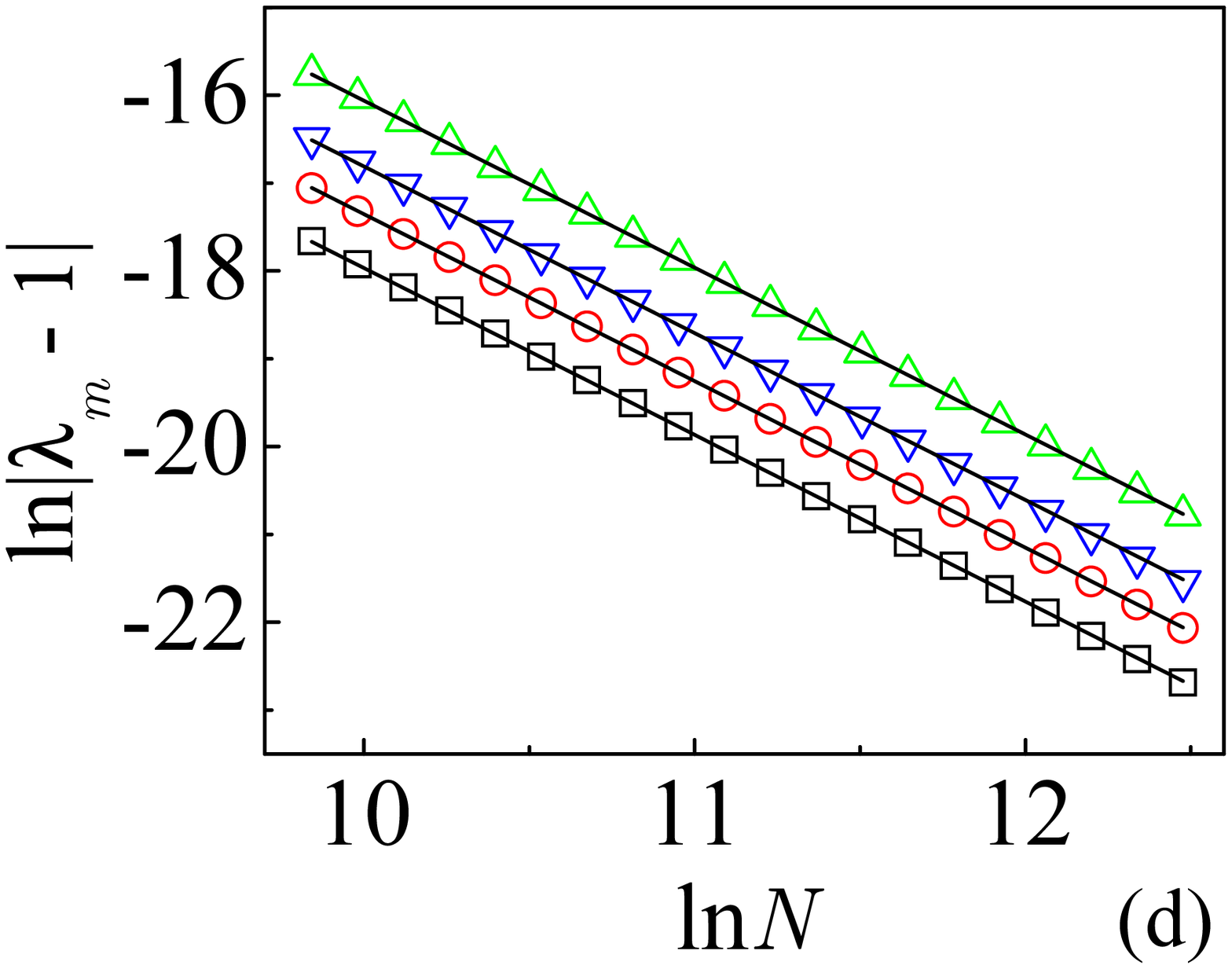}
\caption{(Color online) Scaling behaviors of the susceptibility $\protect%
\chi _{i}(\protect\lambda _{i},\protect\lambda _{f},0)$ with $\protect%
\lambda _{f}=0$, $0.6$, $1.5$\ and $\infty $. (a) and (b) are plots of $%
\protect\chi _{i}(\protect\lambda _{i},\protect\lambda _{f},0)$\ in the
thermodynamic limit. (c) and (d) are plots of $\protect\chi _{i}(\protect%
\lambda _{m},\protect\lambda _{f},0)$\ and $\ln |\protect\lambda _{m}-1|$\
as functions of $\ln N$, where $\protect\lambda _{m}$\ is the
pseudo-critical point. The slopes of the solid lines are $-\protect\pi ^{-1}$
in (b), $\protect\pi ^{-1}$ in (c), and $-1.9$ in (d).}
\label{Scaling}
\end{figure}

%%%%%%%%%%%%%%%%%%%%%%%%%%%%%%%%%%%%%%%%%%%%%%%%%%%%%%%%%%%%%%%%%%%%%%%%%%%%%%%

\section{Scaling behaviors}

We firstly consider the case of zero-temperature initial state with $\lambda
_{i}\sim 1$, where $\rho (\lambda _{i},0,0)=$ $\left\vert 0(\lambda
_{i})\right\rangle \left\langle 0(\lambda _{i})\right\vert $. It should be
noted that in the thermodynamic limit, the ground state of $H(\lambda _{i})$
is two-fold degenerate in the region $\lambda _{i}<1$. Fortunately, any
combination of these two degenerate states as an initial state will give the
same result. From Eq. (\ref{mq}) we have
\begin{equation}
m_{q}(\lambda _{i},\lambda _{f},0)=\left\langle 0(\lambda _{i})\right\vert
\widehat{m}_{q}(\lambda _{f})\left\vert 0(\lambda _{i})\right\rangle ,
\label{mq0}
\end{equation}%
which writes $m_{q}(\lambda _{i},\lambda _{f},0)$ as an expectation value of
the operator $\widehat{m}_{q}(\lambda _{f})$ in the ground state $\left\vert
0(\lambda _{i})\right\rangle $ and presumably implies that a sudden change
of the ground state $\left\vert 0(\lambda _{i})\right\rangle $ around $%
\lambda _{i}=1$ will lead to some critical behaviors of $m_{q}(\lambda
_{i},\lambda _{f},0)$. Actually, in the strong-field limit $\lambda
_{f}=\infty $, the transverse field term is dominant, which keeps the
conservation of the expectation value of the total spin component in the
transverse field direction during the quench process. Consequently, we have $%
m_{q}(\lambda _{i},\infty ,0)=$ $m_{q}(\lambda _{i},\lambda _{i},0)$\ and $%
\chi _{i}(\lambda _{i},\infty ,0)=$ $\chi _{i}(\lambda _{i},\lambda _{i},0)$%
, where $m_{q}(\lambda _{i},\lambda _{i},0)$ and $\chi _{i}(\lambda
_{i},\lambda _{i},0)$\ are the static magnetization and susceptibility,
respectively. $\chi _{i}(\lambda _{i},\infty ,0)$\ should share the same
critical behaviors as those of $\chi _{i}(\lambda _{i},\lambda _{i},0)$,\
the second order derivative of the ground-state energy density. So $\chi
_{i}(\lambda _{i},\infty ,0)$ should diverge at $\lambda _{i}=1$ and obey
scaling behaviors at\ $\lambda _{i}\sim 1$.

Next we will show that a similar conclusion could be obtained for finite $%
\lambda _{f}$. Define $\Lambda _{i}=$ $\min \{\lambda _{i},\lambda
_{i}^{-1}\}$ and $\Lambda _{f}=$ $\min \{\lambda _{f},\lambda _{f}^{-1}\}$.
When $\Lambda _{i}>\Lambda _{f}$, from Eq. (\ref{In}) we have \cite{Integral}
\begin{equation}
m_{q}(\lambda _{i},\lambda _{f},0)=\int_{0}^{\Lambda _{i}}\frac{dx}{x}\left(
\frac{v}{\pi }+w\sqrt{\frac{\Lambda _{i}}{4x}}\right) +w,
\end{equation}%
where $v=$ $\mathrm{sgn}(\Lambda _{f}-x)c(\lambda _{f},x)[s(\lambda
_{f},x)s(\lambda _{i},x)$ $-c(\lambda _{f},x)c(\lambda _{i},x)]$, $c(\lambda
,x)=$ $[2\lambda -(x+x^{-1})]/d$, $s(\lambda ,x)$ $=(x-x^{-1})/d$, $%
d(\lambda ,x)=$ $2\sqrt{|1+\lambda ^{2}-\lambda (x+x^{-1})|}$ and $w=$ $%
(\lambda _{i}+\lambda _{f})/(2\pi \lambda _{f}\sqrt{\lambda _{i}\Lambda _{i}}%
)$. Accordingly, the susceptibility is
\begin{equation}
\chi _{i}(\lambda _{i},\lambda _{f},0)=\int_{0}^{\Lambda _{i}}\frac{dx}{x}%
\frac{s(\lambda _{i},x)}{d(\lambda _{i},x)}\frac{u}{\pi }+C(\lambda
_{i},\lambda _{f}),
\end{equation}%
where $u=$ $\mathrm{sgn}(\Lambda _{f}-x)c(\lambda _{f},x)[s(\lambda
_{f},x)c(\lambda _{i},x)$ $-c(\lambda _{f},x)s(\lambda _{i},x)]$ and $%
C(\lambda _{i},\lambda _{f})$ is convergent at $\lambda _{i}=1$. In the case
of $\Lambda _{i}\simeq 1$, i.e., $\lambda _{i}\sim 1$, the above integral
can be reduced as $\int_{0}^{\Lambda _{i}}(...)dx\simeq $ $\int_{\Lambda
_{i}-\varepsilon }^{\Lambda _{i}}(...)dx$\ with $\varepsilon \ll 1$. Via
straightforward calculations, an asymptotic behavior of $\chi _{i}(\lambda
_{i},\lambda _{f},0)$ around $\lambda _{i}=1$ is obtained as
\begin{equation}
\chi _{i}\left( \lambda _{i},\lambda _{f},0\right) \approx -\frac{1}{\pi }%
\ln |\lambda _{i}-1|+K_{1}(\lambda _{f}),  \label{scaling1}
\end{equation}%
where $K_{1}(\lambda _{f})$\ is a $\lambda _{f}$-dependent constant. In Fig. %
\ref{Scaling} (a) and (b), $\chi _{i}(\lambda _{i},\lambda _{f},0)$ are
plotted as functions of $\lambda _{i}$\ and $\ln |\lambda _{i}-1|$ with $%
\lambda _{f}=0$, $0.6$, $1.5$ and $\infty $.

According to the finite size scaling ansatz \cite{Barber}, the above
critical behavior can be extracted from finite samples, which is important
for quantum simulations. Numerical simulations for finite systems show that
the susceptibility reaches the maximum $\chi _{i}(\lambda _{m},\lambda
_{f},0)$ at the pseudo-critical point $\lambda _{m}$. In\ Fig. \ref{Scaling}
(c), we plot $\chi _{i}(\lambda _{m},\lambda _{f},0)$\ as a function of $\ln
N$ with $\lambda _{f}=0$, $0.6$, $1.5$ and $\infty $. As expected, $\chi
_{i}(\lambda _{m},\lambda _{f},0)$ diverges logarithmically as
\begin{equation}
\chi _{i}(\lambda _{m},\lambda _{f},0)\approx \frac{1}{\pi }\ln
N+K_{2}(\lambda _{f}),  \label{scaling2}
\end{equation}%
where $K_{2}(\lambda _{f})$\ is another $\lambda _{f}$-dependent constant.
As the lattice size approaches infinite, the pseudo-critical point $\lambda
_{m}$ tends to the QCP as $|\lambda _{m}-1|\propto N^{-1.9}$, which is shown
in Fig. \ref{Scaling} (d). According to the scaling ansatz in the case of
logarithmic divergence \cite{Barber}, the ratio between the two prefactors
of the logarithm in Eq. (\ref{scaling1}) and (\ref{scaling2}) is the
exponent $\nu $ that governs the divergence of the correlation length. As
expressed in Eq. (\ref{scaling1}) and (\ref{scaling2}), numerical
calculations give $\nu =1$, which accords with the result obtained from the
exact solution of the TFIM \cite{Barber}.

When the initial system is at finite temperature, the correlation length of $%
\rho (\lambda _{i},T,0)$ is always convergent and all the scaling behaviors
will vanish. We do not discuss this case in detail.

\section{Discontinuity of the quench susceptibility}

Now we focus on $m_{q}(\lambda _{i},\lambda _{f},T)$ and $\chi _{f}(\lambda
_{i},\lambda _{f},T)$ at $\lambda _{f}\sim 1$.\ Mathematically, under the
condition $\Lambda _{f}>\Lambda _{i}$, Eq. (\ref{In}) becomes
\begin{eqnarray}
m_{q}(\lambda _{i},\lambda _{f},T) &=&\frac{|1-\lambda _{f}^{2}|}{4\lambda
_{f}^{2}}\tanh \Theta \lbrack c(\lambda _{i},\Lambda _{f})  \label{mqT} \\
&&-s(\lambda _{i},\Lambda _{f})]+A(\lambda _{i},\lambda _{f},T),  \notag
\end{eqnarray}%
where $\Theta =d(\lambda _{i},\Lambda _{f})/2T$ and $A(\lambda _{i},\lambda
_{f},T)$\ is an analytical function when $\lambda _{i}$ is away from $1$
\cite{Integral}. Obviously, the factor $|1-\lambda _{f}^{2}|$\ in Eq. (\ref%
{mqT}) causes a sudden change of $m_{q}(\lambda _{i},\lambda _{f},T)$ at $%
\lambda _{f}=1$. For two extreme cases, $\lambda _{i}=0$ and $\infty $, Eq. (%
\ref{In}) is completely integrable and could be written as
\begin{equation}
m_{q}(0,\lambda _{f},T)=\tanh (1/T)\left\{
\begin{array}{cc}
\lambda _{f}/2, & \lambda _{f}\leq 1 \\
1/(2\lambda _{f}), & \lambda _{f}>1%
\end{array}%
\right.
\end{equation}%
and
\begin{equation}
m_{q}(\infty ,\lambda _{f},T)=\left\{
\begin{array}{cc}
1/2, & \lambda _{f}\leq 1 \\
1-1/(2\lambda _{f}^{2}), & \lambda _{f}>1%
\end{array}%
\right. .
\end{equation}

Accordingly, such a sudden change leads to a discontinuity of $\chi
_{f}(\lambda _{i},\lambda _{f},T)$ at the QCP. The expression is
\begin{eqnarray}
\chi _{f}(\lambda _{i},\lambda _{f},T) &=&\mathrm{sgn}(\lambda _{f}-1)\frac{%
\tanh \Theta }{2\lambda _{f}^{3}}[c(\lambda _{i},\Lambda _{f}) \\
&&-s(\lambda _{i},\Lambda _{f})]+B(\lambda _{i},\lambda _{f},T),  \notag
\end{eqnarray}%
where $B(\lambda _{i},\lambda _{f},T)$ is continuous at $\lambda _{f}\sim 1$%
. The $\mathrm{sgn}$ function leads to a jump of $\chi _{f}(\lambda
_{i},\lambda _{f},T)$ with magnitude $\delta \chi _{f}=$ $\tanh (\Delta
_{i}/2T)$, where $\Delta _{i}=$ $2|\lambda _{i}-1|$ is the energy gap above
the ground state for the initial Hamiltonian $H(\lambda _{i})$. Remarkably,
the jump $\delta \chi _{f}$ always occurs at the QCP even for a
finite-temperature initial state. When $T\ll \Delta _{i}$, $\delta \chi _{f}$
decays slowly as $1-2\exp (-\Delta _{i}/T)$, and is almost a constant within
the range $T<0.1\Delta _{i}$. In this sense, the lower-temperature samples
share the same feature as that of the zero-temperature sample, which is
crucial for the experimental detection of critical behaviors of QPT systems,
since cooling matter to zero temperature is impossible in any experiment.

%%%%%%%%%%%%%%%%%%%%%%%%%%%%%%%%%%%%%%%%%%%%%%%%%%%%%%%%%%%%%%%%%%%%%%%%%%%%%%%

\begin{figure}[tbp]
\includegraphics[bb=15 325 535 735, width=4 cm]{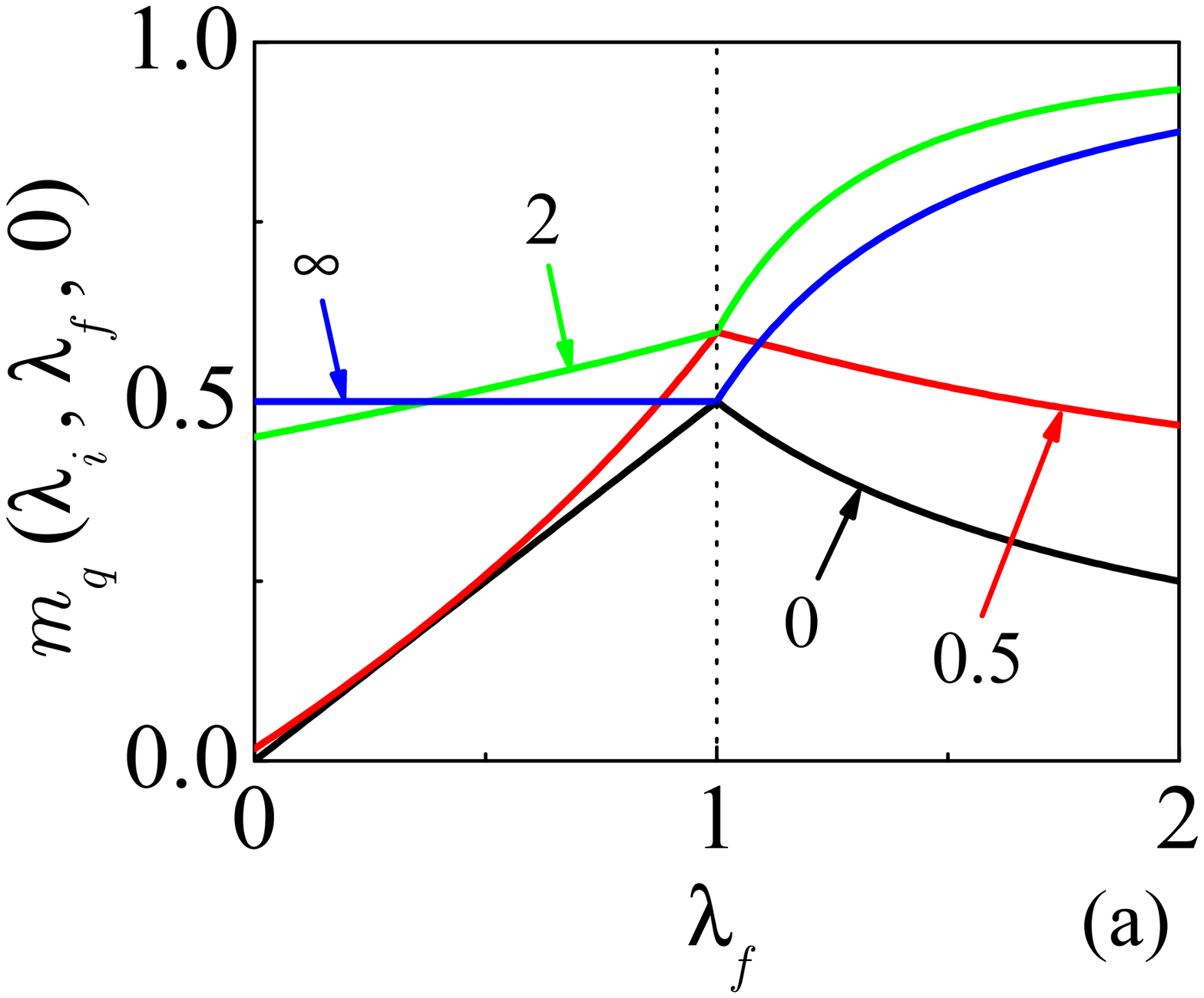} %
\includegraphics[bb=15 325 535 735, width=4 cm]{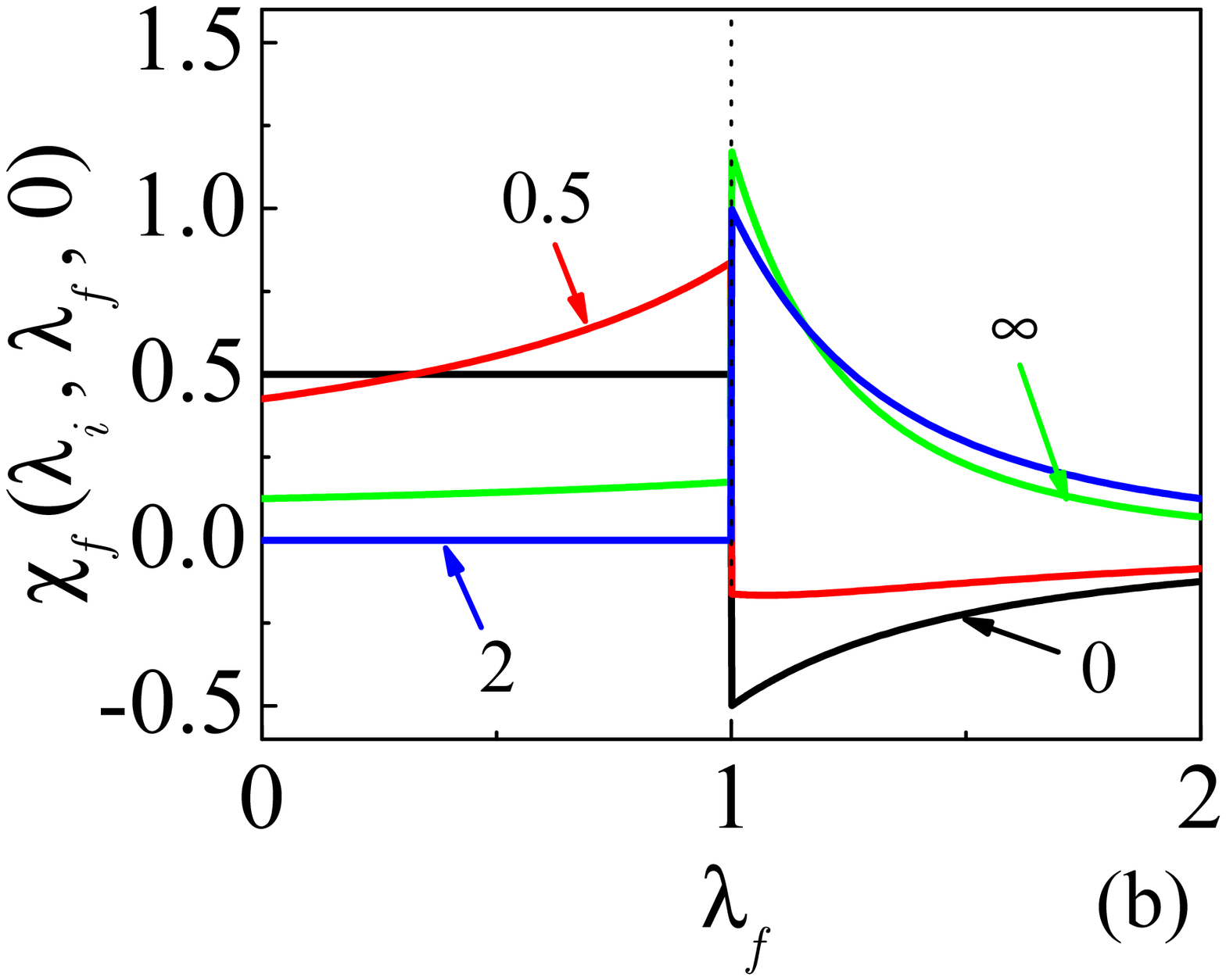}
\caption{(Color online) Plots of $m_{q}(\protect\lambda _{i},\protect\lambda %
_{f},0)$\ and $\protect\chi _{f}(\protect\lambda _{i},\protect\lambda %
_{f},0) $\ with $\protect\lambda _{i}=0$, $0.5$, $2$\ and $\infty $. $m_{q}(%
\protect\lambda _{i},\protect\lambda _{f},0)$\ has a sudden change at\ $%
\protect\lambda _{f}=1$,\ while $\protect\chi _{f}(\protect\lambda _{i},%
\protect\lambda _{f},0)$\ has a discontinuity at $\protect\lambda _{f}=1$.
When $\protect\lambda _{f}$\ passes through $1$,\ $\protect\chi _{f}(\protect%
\lambda _{i},\protect\lambda _{f},0)$ drops down for $\protect\lambda _{i}<1$
and jumps up for $\protect\lambda _{i}>1$.}
\label{Dis}
\end{figure}

%%%%%%%%%%%%%%%%%%%%%%%%%%%%%%%%%%%%%%%%%%%%%%%%%%%%%%%%%%%%%%%%%%%%%%%%%%%%%%%

To exhibit the above critical behaviors, in Fig. \ref{Dis}, $m_{q}(\lambda
_{i},\lambda _{f},0)$ and $\chi _{f}(\lambda _{i},\lambda _{f},0)$ with $%
\lambda _{i}=0$, $0.5$, $2$ and $\infty $ are plotted. It is shown that $%
m_{q}(\lambda _{i},\lambda _{f},0)$ has a sudden change for systems with a
wide range of $\lambda _{i}$. For $\lambda _{f}<1$, $m_{q}(\lambda
_{i},\lambda _{f},0)$ increases as $\lambda _{f}$ increases, except that
when $\lambda _{i}=\infty $, $m_{q}(\lambda _{i},\lambda _{f},0)$ is always
equal to $0.5$. In contrast, for $\lambda _{f}>1$, $m_{q}(\lambda
_{i},\lambda _{f},0)$ increases for $\lambda _{i}>1$ and decreases for $%
\lambda _{i}<1$. Accordingly, $\chi _{f}(\lambda _{i},\lambda _{f},0)$\ has
a discontinuity at $\lambda _{f}=1$ in the following manner: it drops down
for $\lambda _{i}<1$ and jumps up for $\lambda _{i}>1$.

It is well known that the ground state of the system $H(\lambda _{f})$\
experiences a sudden change when $\lambda _{f}$\ passes through the QCP. A
natural question is whether the sudden change of the ground state causes the
critical behaviors of quench quantities directly. Consider the simplest case
with zero-temperature initial state, where the jump $\delta \chi _{f}\ $gets
the maximum $1$. From Eq. (\ref{mq0}), the contribution of the final ground
state $\left\vert 0(\lambda _{f})\right\rangle $\ to the quench quantity is
proportional to the fidelity of two ground states \cite{GuSJ,ChenShu}
\begin{equation}
\left\vert \langle 0(\lambda _{i})\left\vert 0(\lambda _{f})\right\rangle
\right\vert ^{2}=\prod_{k}\cos \frac{\theta _{k}(\lambda _{f})-\theta
_{k}(\lambda _{i})}{2}.
\end{equation}%
Straightforward calculations show that $\left\vert \langle 0(\lambda
_{i})\left\vert 0(\lambda _{f})\right\rangle \right\vert $\ is always
vanishing for finite $|\lambda _{f}-\lambda _{i}|$ in the thermodynamic
limit. Thus the critical behaviors of quench quantities are not direct
consequences of the sudden charge of the ground state, which indicates that
the excited states also experience drastic changes at the QCP.

%%%%%%%%%%%%%%%%%%%%%%%%%%%%%%%%%%%%%%%%%%%%%%%%%%%%%%%%%%%%%%%%%%%%%%%%%%%%%%%

\begin{figure}[tbp]
\includegraphics[bb=3 312 546 765, width=4 cm]{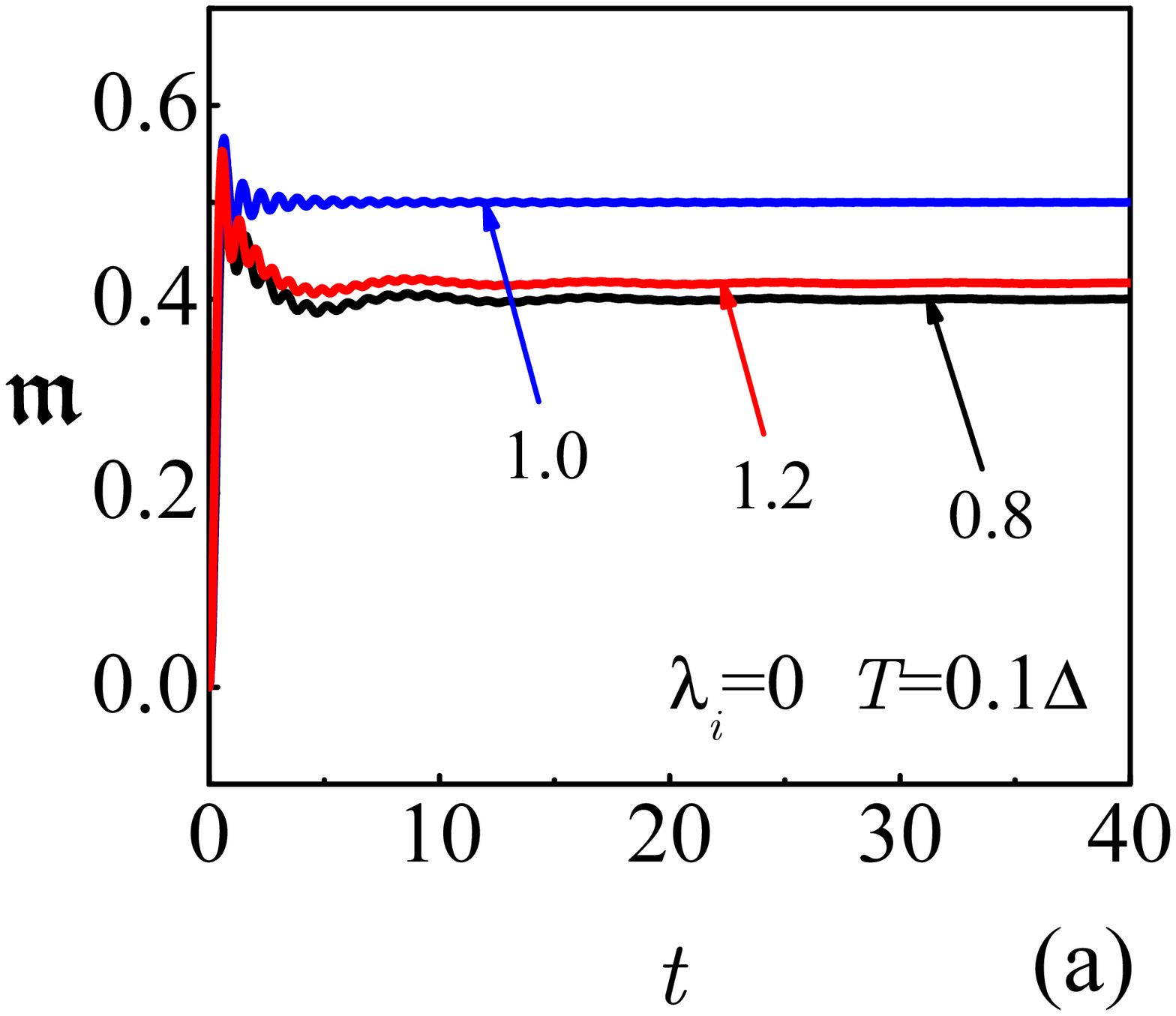} %
\includegraphics[bb=3 312 546 765, width=4 cm]{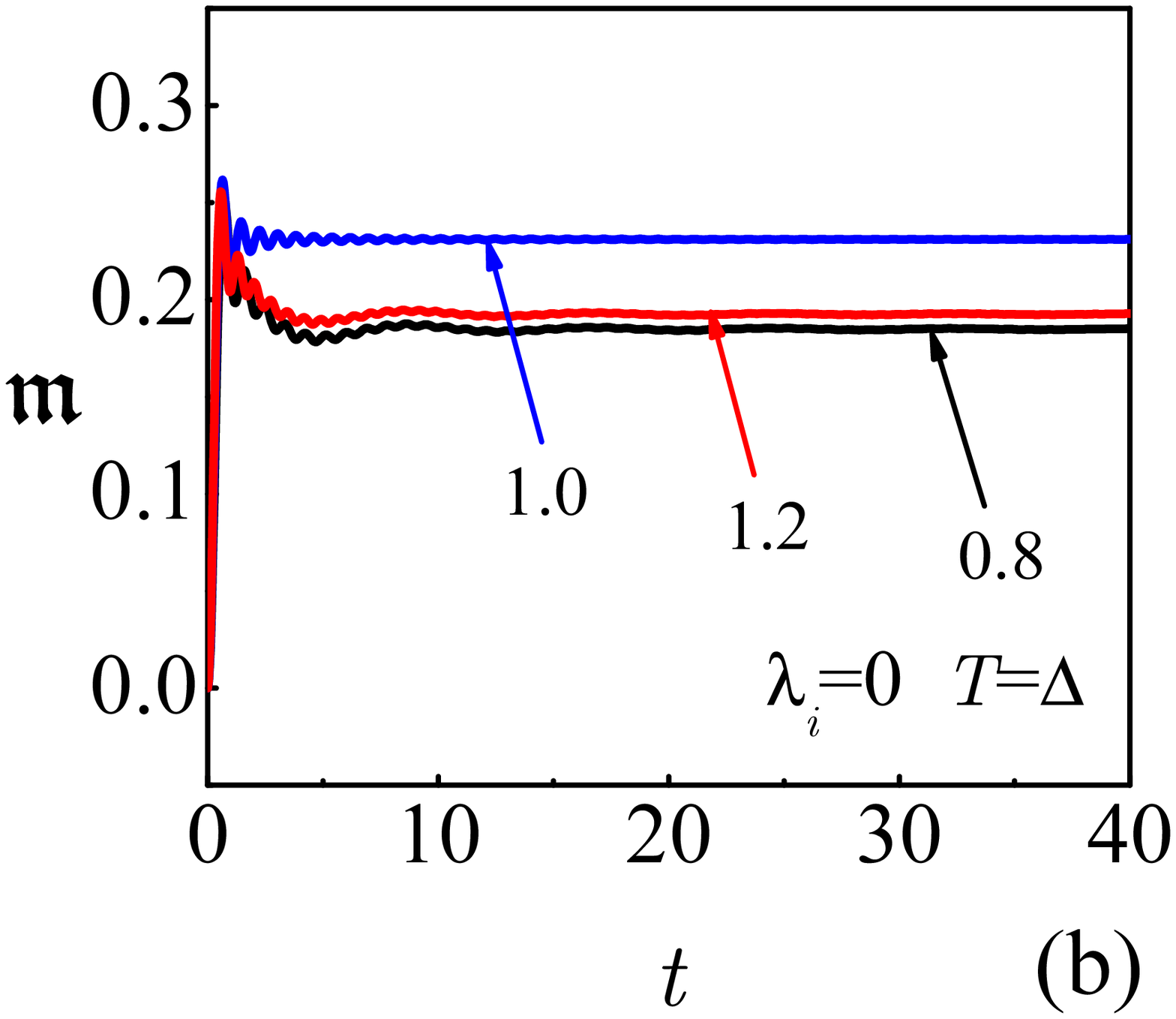} %
\includegraphics[bb=3 312 546 765, width=4 cm]{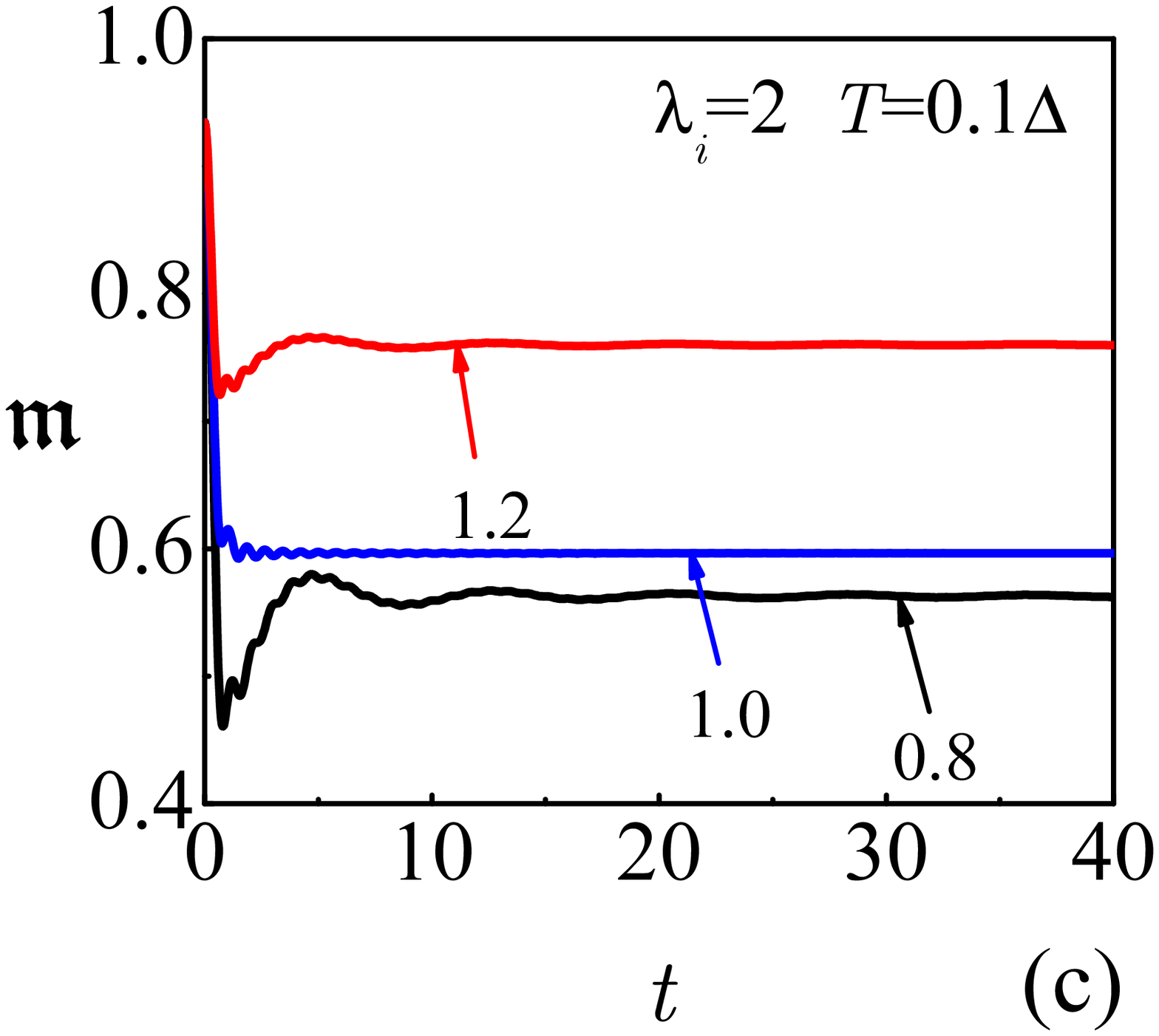} %
\includegraphics[bb=3 312 546 765, width=4 cm]{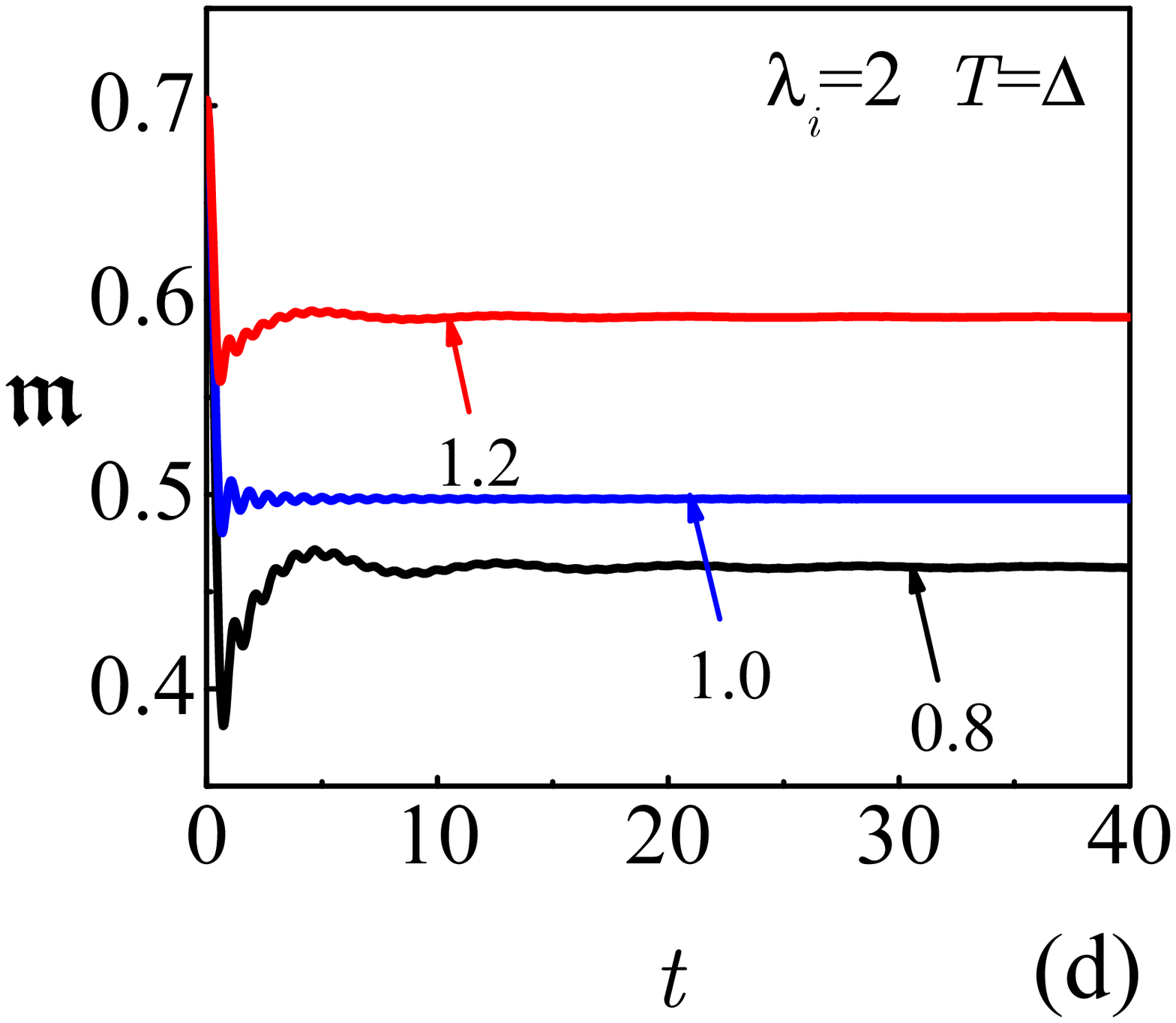}
\caption{(Color online) The time evolution of $\mathfrak{m}(\protect\lambda %
_{i},\protect\lambda _{f},T,t)$\ with $\protect\lambda _{i}=0$, $2$; $%
T/\Delta _{i}=0.1$, $1$\ and $\protect\lambda _{f}=0.8$, $1$, $1.2$. $%
\mathfrak{m}(\protect\lambda _{i},\protect\lambda _{f},T,t)$\ oscillates
initially and becomes steady at $t=15$\ approximately.}
\label{TiEv}
\end{figure}

%%%%%%%%%%%%%%%%%%%%%%%%%%%%%%%%%%%%%%%%%%%%%%%%%%%%%%%%%%%%%%%%%%%%%%%%%%%%%%%

\section{Discussion and conclusion}

In above analyses, we neglect the interaction between the sample of TFIM and
its environment. Generally, such an interaction will induce decoherence of
the system. Therefore, our results are valid only when the quench relaxation
time $\tau _{Q}$ is short enough compared to the decoherence time. In order
to estimate the order of $\tau _{Q}$,\ numerical simulations for $\mathfrak{m%
}(\lambda _{i},\lambda _{f},T;t)$\ are performed. In Fig. \ref{TiEv}, we
plot the results with $\lambda _{i}=0,2$, $T/\Delta _{i}=0.1,1$ and $\lambda
_{f}=0.8,1,1.2$ as examples. Remarkably, $\tau _{Q}$ is not sensitive to $T$%
, $\lambda _{i}$ and $\lambda _{f}$, and $\tau _{Q}$ is smaller than $%
15J^{-1}$, where $J$ is the Ising coupling strength and in this paper $J=1$.
In a spin network for quantum information processing (QIP), the shortest
period of time for an operation between two neighbor qubits via a natural
dynamics is $(2J)^{-1}$. Therefore the realization of a complete quench
process is not a difficult task for a spin network which is utilized for QIP
in practice.

In summary, we found that the quench magnetic susceptibility as a function
of the initial field strength exhibits strongly similar scaling behaviors to
those of the adiabatic process, and the quench magnetic susceptibility as a
function of the final field strength shows a discontinuity at the QCP, which
remains robust even when the initial system is at finite temperature. This
observation is useful for understanding QPTs and studying the properties of
the QPT systems. Moreover, it gives us a new approach to observe the QCP and
critical behaviors using low-temperature samples experimentally, avoiding
the rigorous restriction of zero temperature. Because of the universality
principle, the critical behaviors only depend on the dimension and the
breaking symmetry, so the obtained results are heuristic and may be extended
to other QPT models.

\begin{acknowledgments}
This work was supported by the CNSF with Grants No. 10874091 and No.
2006CB921205.
\end{acknowledgments}


\begin{thebibliography}{99}
\bibitem{Integral} The integral in Eq. (\ref{In}) can be written as a
circular integral in a complex plane via the transformation $z=e^{ik}$. For $%
\Lambda _{f}>\Lambda _{i}$, there is a branch cut along the real axis from $%
0 $ to $\Lambda _{i}$. For $\Lambda _{f}<\Lambda _{i}$, besides the branch
cut, there is an isolated singularity at $z=\Lambda _{f}$.
\end{thebibliography}
\end{document}